\documentclass[runningheads]{llncs}

\usepackage{amssymb}
\usepackage{subfigure}
\usepackage{graphicx}
\usepackage{graphics}
\usepackage{colortbl}
\usepackage{times}
\usepackage{helvet}
\usepackage{courier}
\usepackage{verbatim}
\usepackage{url}
\usepackage{epstopdf}
\usepackage{algorithm}
\usepackage{algorithmic}
\usepackage{float}
\usepackage{makecell}
\usepackage{multirow}
\usepackage{textcomp}
\usepackage{float}
\usepackage{bm}
\usepackage{hyphenat}
\usepackage{colortbl}
\usepackage{caption}
\usepackage{mdwlist}
\usepackage{CJK}
\usepackage{amsmath}
\usepackage{caption}
\usepackage{subfigure}

\urldef{\mails}\path|yzheng66@iit.edu|

\newcommand{\keywords}[1]{\par\addvspace\baselineskip
\noindent\keywordname\enspace\ignorespaces#1}

\begin{document}

\mainmatter  

\title{A User's Guide to CARSKit}

\author{Yong Zheng}

\institute{DePaul University, Chicago, IL, USA 60604\\
Illinois Institute of Technology, Chicago, IL, USA 60616\\
\mails\\
}

\maketitle

\begin{abstract}
Context-aware recommender systems extend traditional recommenders by adapting their suggestions to users' contextual situations. CARSKit is a Java-based open-source library specifically designed for the context-aware recommendation, where the state-of-the-art context-aware recommendation algorithms have been implemented. This report provides the basic user's guide to CARSKit, including how to prepare the data set, how to configure the experimental settings, and how to evaluate the algorithms, as well as interpreting the outputs. The instructions in this guide are applicable for CARSKit v0.3.5 and above.

\end{abstract}


\section{Introduction}
\noindent
CARSKit~\cite{zheng2015carskit}~\footnote{https://github.com/irecsys/CARSKit/} is an open-source Java-based context-aware recommendation engine, where it can be used, modified and distributed under the terms of the GNU General Public License. (Java version 1.7 or higher required). It is specifically designed for context-aware recommendations.

CARSKit is a free software: you can redistribute it and/or modify it under the terms of the GNU General Public License (GPL) as published by the Free Software Foundation, either version 3 of the License, or (at your option) any later version. CARSKit is distributed in the hope that it will be useful, but WITHOUT ANY WARRANTY; without even the implied warranty of MERCHANTABILITY or FITNESS FOR A PARTICULAR PURPOSE. See the GNU General Public License for more details. You should have received a copy of the GNU General Public License along with CARSKit. If not, see http://www.gnu.org/licenses/.

\subsection{Design}
\begin{figure*}[htbp]
\setlength{\abovecaptionskip}{1pt}
\centering
\includegraphics[scale=0.60]{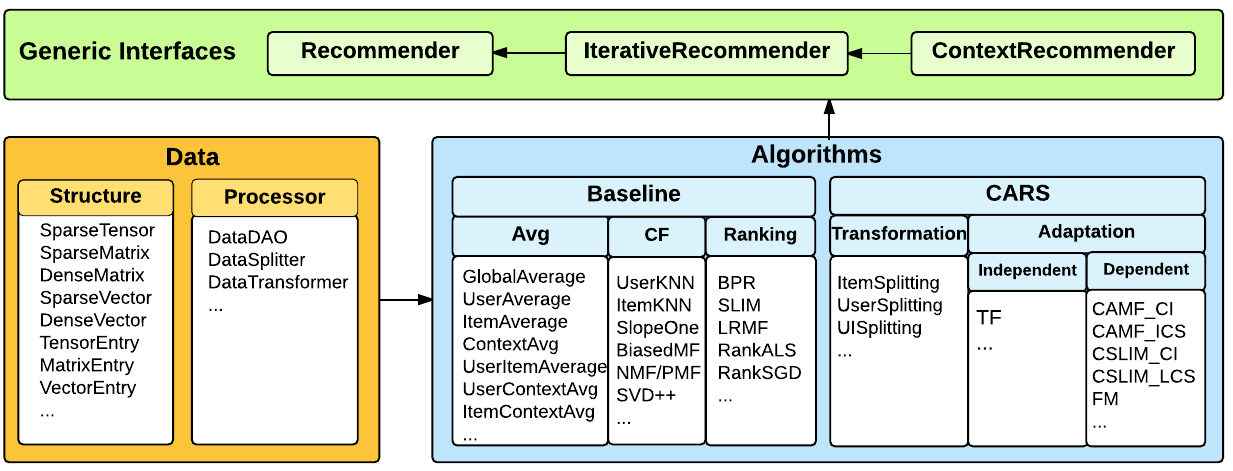}
\caption{CARSKit: The Architecture and Design}
\label{fig:design}
\end{figure*}

\noindent
CARSKit provides a flexible architecture so that it is easy to expand the scope of context-aware recommendation algorithms and provides spaces to develop new algorithms in the future. The whole design can be depicted by the Figure~\ref{fig:design}.

The workflow is straightforward in our design: different recommendation algorithms are the specific implementations and extensions from the generic interfaces where the shared and common functions are defined, such as rating or score prediction for a user on one item in a specific context. Evaluations for rating predictions and top-$N$ recommendations are embedded into the $Recommender$.

\subsection{Algorithms}
\noindent
We divide the contextual algorithms into two categories in CARSKit: transformation algorithms and adaptation algorithms.

The transformation algorithms try to pre-process the data and convert the contextual data set to a 2-dimensional rating matrix which only contains users, items and ratings, so that any traditional recommendation algorithms can be applied to. One of those techniques is the context-aware splitting approaches~\cite{baltrunas2013experimental,zheng2014splitting}.

The adaptation algorithms directly incorporate contexts into the prediction function. There are two subcategories involved: \textit{independent modeling} (e.g., TF~\cite{karatzoglou2010multiverse}) which assumes contexts are independent with users (and items), and \textit{dependent modeling} which exploits the dependencies among users, items and contexts, such as CAMF~\cite{baltrunas2011matrix} and contextual sparse linear method (CSLIM)~\cite{zheng2014cslim,zheng2014cikm}. Dependent modeling can be built in two ways: by modeling contextual rating deviations~\cite{baltrunas2011matrix,zheng2014cikm} and by learning context similarities~\cite{zheng2015umap,zheng2015simcars}. Factorization machines (FM)~\cite{rendle2011fast} is a finer-grained algorithm which exploits pairwise relationships in its learning process. Among those algorithms, TF and CAMF are two popular ones which have been recognized as the standard baselines in CARS.

In addition to those state-of-the-art contextual recommendation algorithms, we also include some traditional recommendation algorithms in the package \textit{baseline}. We did not re-compile those algorithms and directly reuse the implementations in LibRec~\footnote{https://github.com/guoguibing/librec/}. There are two main purposes to include those traditional recommendation algorithms -- On one hand, those algorithms can be applied after the data transformation (e.g., splitting operations), which is an essential step in the context-aware transformation algorithms. On the other hand, it is usually common to compete a contextual recommendation algorithm with non-contextual algorithms to judge whether the contextual effect is significant or a context-aware recommendation algorithm is necessary or not.

\subsection{Evaluations}
\noindent
Most of the algorithms embedded in CARSKit are able to perform the two recommendation task: \textit{rating prediction} and \textit{item recommendation}, except those ones specifically designed for top-$N$ recommendation, such as CSLIM. But the evaluation is different from traditional ones, since contexts are additional inputs in the evaluation process. Typically, the rating prediction can be evaluated by different prediction errors, such as mean absolute error (MAE), root mean square error (RMSE) and mean prediction error (MPE). The item recommendation can be evaluated through \textit{relevance} metrics, such as precision and recall, and \textit{ranking} metrics, such as mean average precision (MAP), normalized discounted cumulative gain (NDCG) and mean reciprocal rank (MMR).

\section{User's Guide}
\noindent
In this section, the specific instructions about how to use, deploy and evaluate the context-aware recommendation algorithms by CARSKit are introduced as follows.

\subsection{Data Format}
\noindent
Usually, the contextual rating data can be stored in two formats: \textit{loose} format and \textit{compact} format, as shown in tables below. Note that, the names of the last two columns in the \textit{loose format} must be ``dimension" and ``condition".

\vspace{-15pt}
\begin{table}
\caption{Data Format}\label{tab:dataformat}
\centering\scriptsize
\subtable[Loose Format]{
      \begin{tabular}{|c|c|c|c|c|}
\hline
{\bf UserID} & {\bf ItemID} & {\bf Rating} & {\bf Dimension} & {\bf Condition} \\ \hline
U1           & T1           & 3            & Time          & Weekend         \\ \hline
U1           & T1           & 3            & Location      & Work            \\ \hline
U2           & T2           & 4            & Time          & Weekday         \\ \hline
U2           & T2           & 4            & Location      & Home            \\ \hline
\end{tabular}
}
\qquad
\subtable[Compact Format]{
\begin{tabular}{|c|c|c|c|c|}
\hline
{\bf UserID} & {\bf ItemID} & {\bf Rating} & {\bf Time} & {\bf Location} \\ \hline
U1           & T1           & 3            & Weekend    & Work           \\ \hline
U2           & T2           & 4            & Weekday    & Home           \\ \hline
U1           & T1           & 4            & Weekend    & Home           \\ \hline
U2           & T2           & 2            & Weekday    & Work           \\ \hline
\end{tabular}
}
\end{table}
\vspace{-12pt}

Context dimension is identical to contextual variable, e.g., \textit{Time} and \textit{Location} as shown in the example above. Context conditions refer to specific values in a dimension, e.g. \textit{Weekend} and \textit{Weekday} are two conditions in the dimension \textit{Time}. The loose format assumes that there is only one rating for each $<$user, item$>$ pair in associated contexts, where the compact format allows to store multiple ratings to a same $<$user, item$>$ pair in different contextual situations. Take the example shown in the two tables above, the first two rows in loose format actually represent a single rating by U1 for T1 within contexts \{Weekend, Work\}. In compact format, each row represents a single contextual rating profile; that is, there are only two contextual rating profiles in the loose format but four rating profiles in the compact format in this example.

\vspace{-12pt}
\begin{table}[ht!]\scriptsize
\centering
\tabcolsep 0.8pt
\begin{tabular}{|c|c|c|c|c|c|c|}
\hline
{\bf UserID} & {\bf ItemID} & {\bf Rating} & {\bf Time:Weekend} & {\bf Time:Weekday} & {\bf Location:Home} & {\bf Location:Work} \\ \hline
U1           & T1           & 3            & 1                  & 0                  & 0                   & 1                   \\ \hline
U2           & T2           & 4            & 0                  & 1                  & 1                   & 0                   \\ \hline
U1           & T1           & 4            & 1                  & 0                  & 1                   & 0                   \\ \hline
U2           & T2           & 2            & 0                  & 1                  & 0                   & 1                   \\ \hline
\end{tabular}
\caption{Binary Format}
\label{tab:format}
\end{table}
\vspace{-16pt}

Most contextual information is in shape of categorical data. Both the loose and compact format will increase storage pressure and computational costs. In CARSKit, we store contextual rating in a \textit{binary} format as shown in Table~\ref{tab:format}, which is able to significantly boost the running performance. To assist the end users to prepare the rating data, we provide two methods \textit{TransformationFromLooseToBinary} and \textit{TransformationFromCompactToBinary} as the data transformer in our toolkit.

\textbf{Special Notes: } If you do have missing values in the context conditions, just leave it as empty entries, or simply set ``na" (NA, Not Available) as the value of the context conditions. This setting is useful for two scenarios -- either you have missing values in context conditions, or you blend ratings with \& without considering contexts.

\subsection{Data Preparation}
\noindent
It is better to follow the steps below to prepare your data:
\begin{itemize}
    \item \textbf{Step 1. }Prepare your data set in either Loose or Compact format, and the CARSKit will automatically convert your data to the binary format. Be advised that you should add header to your data as shown in the Table~\label{ref}. Save your file to either txt or csv format by using comma as separator.
    \item \textbf{Step 2. }Create a folder for each data set, and put the ratings.txt or ratings.csv in each folder. It is suggested, since the CARSKit will create a working subfolder under the path where your rating data is stored.
    \item \textbf{Step 3. }Assign the data path to the configuration file and run any algorithms to test the data transformation.
\end{itemize}

\subsection{Experimental Configuration}
\noindent
You are required to create a setting.conf figure which is already included in the CARSKit repository~\footnote{https://github.com/irecsys/CARSKit/}. The main pieces to be configured in this file can be introduced as follows:

\subsubsection{Data Path}:\\
\textit{dataset.ratings.wins=C:\textbackslash\textbackslash Data\textbackslash\textbackslash DePaulMovie\textbackslash\textbackslash ratings.txt\\
dataset.ratings.lins=/data/DePaulMovie/ratings.txt}\\
Note: you may set up the data path either for windows (i.e., wins) or non-windows platforms (i.e., lins for Linus/Mac systems)

\subsubsection{Setup Your Ratings}:\\
\textit{ratings.setup=-threshold -1 -datatransformation 1 -fullstat -1}\\
Note: By setting a rating threshold, such as \textit{-threshold 3}, the relevant items in top-$N$ recommendation are defined as the items which were assigned a rating no less than 3. Setting a positive value to \textit{-datatransformation} will help you automatically transform your data to the binary format. Set it as -1 if you do not need data transformations. However, you need to put the train.csv (and/or test.csv) in the folder ``CARSKit.Workspace", if not data transformation is required. CARSKit can output a comprehensive descriptive statistics of your data set if you set a positive value to \textit{-fullstat}.

\subsubsection{Choose An Algorithm}:\\
\textit{recommender=xxx}\\
Note: the list of options can be summarized as follows.
\begin{itemize}
    \item Baseline-Avg recommender: GlobalAvg, UserAvg, ItemAvg, UserItemAvg
    \item Baseline-Context average recommender: ContextAvg, ItemContextAvg, UserContextAvg
    \item Baseline-CF recommender: ItemKNN, UserKNN, SlopeOne, PMF, BPMF, BiasedMF, NMF, SVD++
    \item Baseline-Top-N ranking recommender: SLIM, BPR, RankALS, RankSGD, LRMF
    \item CARS - splitting approaches: UserSplitting, ItemSplitting, UISplitting; algorithm options: e.g., usersplitting -traditional biasedmf -minlength 2
    \item CARS - independent models: CPTF
    \item CARS - dependent-dev models: CAMF\_CI, CAMF\_CU, CAMF\_C, CAMF\_CUCI, CSLIM\_C, CSLIM\_CI, CSLIM\_CU, CSLIM\_CUCI, GCSLIM\_CC
    \item CARS - dependent-sim models: CAMF\_ICS, CAMF\_LCS, CAMF\_MCS, CSLIM\_ICS, CSLIM\_LCS, CSLIM\_MCS, GCSLIM\_ICS, GCSLIM\_LCS, GCSLIM\_MCS
\end{itemize}

Note that the SLIM-based models and the dependent-sim models can only be used for top-$N$ recommendations.

\subsubsection{Setup Recommendation Task}:\\
\textit{item.ranking=off -topN 10}\\
Note: it will be a rating prediction task if your turn off the item.ranking; otherwise, it will run as a top-$N$ recommendation task. By default, CARSKit will produce results in top-5 and top-10 recommendations. If you set a different $N$, CARSKit will additionally produce results based on the predefined value of $N$.

\subsubsection{Evaluation Protocols}:\\
\textit{evaluation.setup=cv -k 5 -p on --rand-seed 1 --test-view all --early-stop RMSE}\\
\textit{evaluation.setup=given-ratio -r 0.8}\\
\textit{evaluation.setup=test-set -f dataPath --test-view all --early-stop RMSE}\\
The three lines above provides examples of evaluations in three scenarios -- K-fold cross validation, 80\% as training, and manual supply of test set.

\subsubsection{Config Your Outputs}:\\
\textit{output.setup=-folder CARSKit.Workspace -verbose on, off --to-clipboard --to-file results.txt}\\
Note: Basically, all the outputs will be put into the folder named as ``CARSKit.Workspace" which is created under your data path. And all the evaluation results will be appended to the same file ``results.txt".

\subsubsection{Parameter Configurations}:\\
Then you are allowed to config the algorithm parameters as shown in the setting.conf

\subsubsection{Run The Toolkit}:\\
\textit{java -jar CARSKit.jar -c setting.conf}\\
\textit{java -jar CARSKit.jar -c CAMF.conf PMF.conf UserSplitting.conf}\\
Note: You are able to run multiple algorithms or configurations by add configuration files to the command above.

\subsection{Output Interpretation}
\begin{figure*}[ht!]
\setlength{\abovecaptionskip}{1pt}
\centering
\includegraphics[scale=0.60]{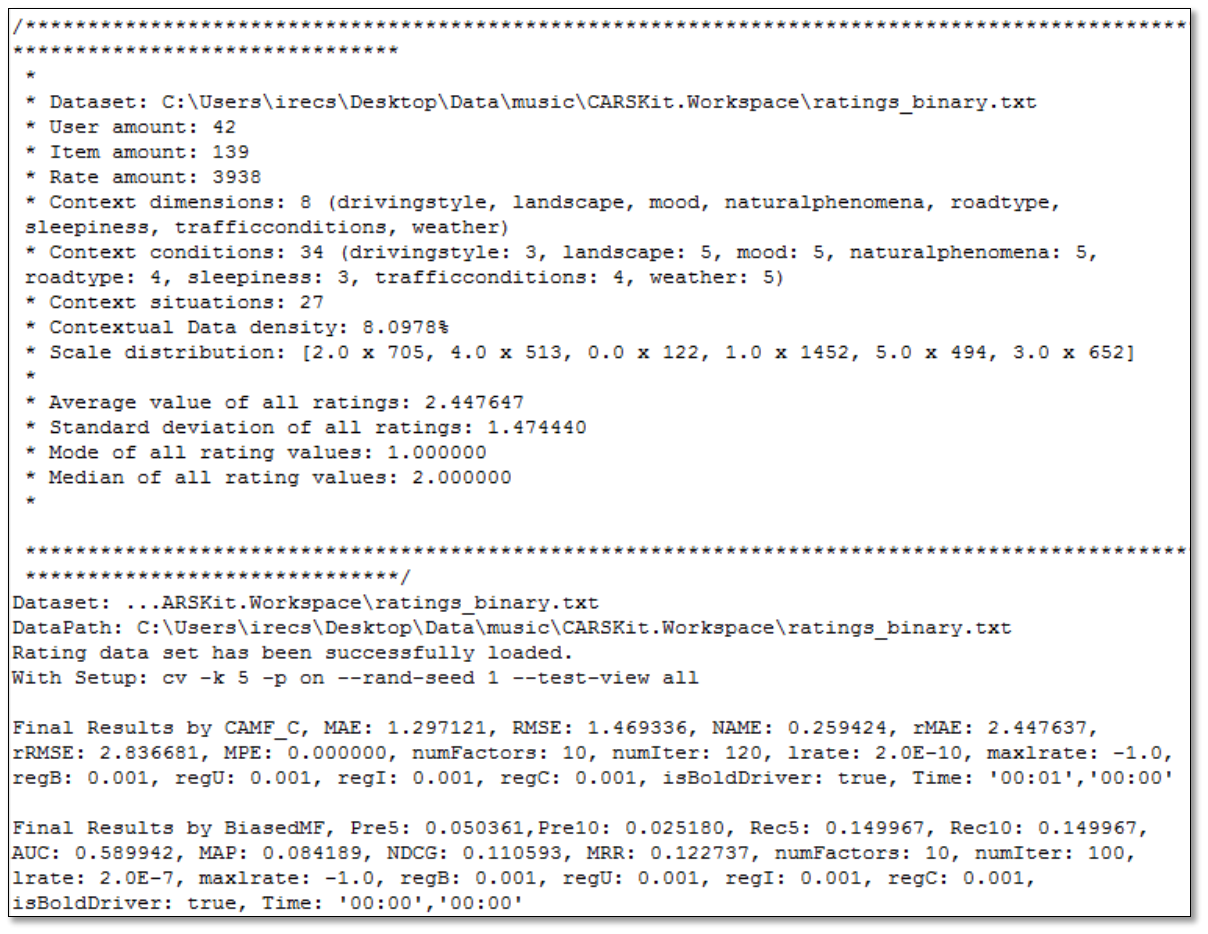}
\caption{Sample of Running Outputs}
\label{fig:output}
\end{figure*}

\noindent
Figure~\ref{fig:output} gives an example of the running outputs.

First of all, the outputs provide a list of simple statistics about the data, such as how many unique users, items and context dimensions or conditions, as well as rating statistics, such as the mean, mode and median in ratings.

In this example, we have results for both rating prediction task (evaluated by prediction errors) and top-N recommendation task (evaluated by precision, recall, MAP, NDCG and MRR) followed by the running parameters so that the users are able to find out the best configurations.

\section{Acknowledgement}
\noindent
I would like to show our gratitude to Dr. Guibing Guo (the author of LibRec) for his comments and suggestions on the development of CARSKit.


\bibliographystyle{abbrv}\scriptsize
\bibliography{bibliography}\scriptsize

\begin{thebibliography}{10}

\bibitem{baltrunas2011matrix}
L.~Baltrunas, B.~Ludwig, and F.~Ricci.
\newblock Matrix factorization techniques for context aware recommendation.
\newblock In {\em Proceedings of the fifth ACM conference on Recommender
  systems}, pages 301--304. ACM, 2011.

\bibitem{baltrunas2013experimental}
L.~Baltrunas and F.~Ricci.
\newblock Experimental evaluation of context-dependent collaborative filtering
  using item splitting.
\newblock {\em User Modeling and User-Adapted Interaction}, pages 1--28, 2013.

\bibitem{karatzoglou2010multiverse}
A.~Karatzoglou, X.~Amatriain, L.~Baltrunas, and N.~Oliver.
\newblock Multiverse recommendation: n-dimensional tensor factorization for
  context-aware collaborative filtering.
\newblock In {\em Proceedings of the fourth ACM conference on Recommender
  systems}, pages 79--86. ACM, 2010.

\bibitem{rendle2011fast}
S.~Rendle, Z.~Gantner, C.~Freudenthaler, and L.~Schmidt-Thieme.
\newblock Fast context-aware recommendations with factorization machines.
\newblock In {\em Proceedings of the 34th international ACM SIGIR conference on
  Research and development in Information Retrieval}, pages 635--644. ACM,
  2011.

\bibitem{zheng2014splitting}
Y.~Zheng, R.~Burke, and B.~Mobasher.
\newblock Splitting approaches for context-aware recommendation: An empirical
  study.
\newblock In {\em Proceedings of the 29th ACM Symposium on Applied Computing},
  pages 274--279. ACM, 2014.

\bibitem{zheng2014cslim}
Y.~Zheng, B.~Mobasher, and R.~Burke.
\newblock {CSLIM}: Contextual {SLIM} recommendation algorithms.
\newblock In {\em Proceedings of the 8th ACM Conference on Recommender
  Systems}, pages 301--304. ACM, 2014.

\bibitem{zheng2014cikm}
Y.~Zheng, B.~Mobasher, and R.~Burke.
\newblock Deviation-based contextual {SLIM} recommenders.
\newblock In {\em Proceedings of the 23rd ACM Conference on Information and
  Knowledge Management}, pages 271--280, 2014.

\bibitem{zheng2015carskit}
Y.~Zheng, B.~Mobasher, and R.~Burke.
\newblock Carskit: A java-based context-aware recommendation engine.
\newblock In {\em Proceedings of the 15th IEEE International Conference on Data
  Mining Workshops}. IEEE, 2015.

\bibitem{zheng2015umap}
Y.~Zheng, B.~Mobasher, and R.~Burke.
\newblock Integrating context similarity with sparse linear recommendation
  model.
\newblock In {\em User Modeling, Adaptation, and Personalization}, pages
  370--376. Springer, 2015.

\bibitem{zheng2015simcars}
Y.~Zheng, B.~Mobasher, and R.~Burke.
\newblock Similarity-based context-aware recommendation.
\newblock In {\em Web Information Systems Engineering}. Springer, 2015.

\end{thebibliography}

\end{document}